\documentclass[12pt]{article}
\usepackage{a4}
\usepackage{supercite}
\usepackage{texdraw}
\usepackage{graphicx}
\def\comment#1{}
\def\y{\zeta}

\def\ev{v}

\def\kxi{(\hbar k\xi)}

\title{Exact Pair Production Rate for a Smooth Potential Step}
\date{}
\author{A. Chervyakov%
\footnote{chervyak@physik.fu-berlin.de~,
on leave from JINR, Dubna, Russia}
\, and
H.~Kleinert%
\footnote{kleinert@physik.fu-berlin.de~,
http://www.physik.fu-berlin.de/\~{}kleinert}
~\\
Institut f\"ur Theoretische Physik,\\
Freie Universit\"at Berlin,\\
Arnimallee 14, D-14195 Berlin}

\begin{document}
\maketitle

\begin{abstract}
We derive the exact rate of pair production of oppositely charged scalar particles
by a smooth potential step $\phi({\bf x})\propto \tanh kz$ in three dimensions. As a check we recover from this the known results
for an infinitely sharp step as well as for a uniform electric field.
\end{abstract}

\section{Introduction}
Spontaneous pair production of oppositely charged particles was first discussed by Heisenberg and Euler~\cite{heeu}
by making use of the Dirac picture of the vacuum. It also provided the solution of the Klein paradox of relativistic quantum mechanics~\cite{klein}
for the scattering on a potential step. Within second quantized field theory it was first discussed by Hund~\cite{hund}, as a precursor
of the later famous calculation of Schwinger~\cite{schwinger}, whose generalization to gravitational fields was made by Hawking~\cite{hawking}.
Schwinger calculated the one-loop effective action of QED in a constant electromagnetic field by the proper time method.
The imaginary part of this effective action yields directly the probability of pair production
 from the vacuum. The result
confirms the fact anticipated by Sauter~\cite{sauter} twenty years earlier that particles can pass through strong repulsive potentials without
the exponential damping  expected in quantum tunneling processes.
A detailed review and the relevant references can be found in Refs.~\cite{caloger} and~\cite{hansen}.

Another method to describe the problem of pair production is due to \cite{nikishov1,nikishov2}.
It is based on the use of causal Green functions~\cite{feynman} to relate the particle production directly to the scattering process.
It can be shown that the rate of pair production can be expressed as an ordinary energy-momentum integral over the logarithm of the reflection coefficient. This formula allows one to  connect in a most transparent and efficient way the one-particle Dirac approach in which the Klein paradox
first appeared with the second quantized field theory in which the problem was first solved satisfactorily.
It has been used for many other developments
including semiclassical approximations~\cite{popov,kim,kl}.
In the same papers~\cite{nikishov1,nikishov2}, also the scattering and pair production processes by the
Sauter potential
$\phi({\bf x})\propto \tanh kz$ were analyzed in detail. The exact solutions of the both Dirac and Klein-Gordon equations were found
in this potential, and the causal Green functions were constructed to
define the pair production
rates of fermions and bosons. However, the calculations were never carried on to derive an exact expression for the
pair production rate. It is the purpose of this paper to complete this gap.

\section{Barrier scattering}
Consider a relativistic scalar particle of charge $e>0$ and mass $m$ moving in an external
electromagnetic potential $A^{\mu} (z) = (\phi (z), \bf 0)$ corresponding to nonuniform electric field along $z$-direction with the strength $E (z) = - \partial\phi (z)/\partial z$. Its kinetic energy and momentum are
\begin{eqnarray}
P_{0} (z) = p_0 - e\phi (z),\quad p_3 (z) = \sqrt{P^{2}_{0} (z) - (p^{2}_{\perp} + m^2)},\quad
p^{2}_{\perp} \equiv p^2_1 + p^2_2.\label{1.1}
\end{eqnarray}
In the transverse direction the particle
propagates freely as a plane wave $\exp[ i({\bf p_\perp}\cdot{\bf x_\perp} - p_0 \,x_0)/\hbar]$.
Thus we represent the solution of Klein-Gordon equation in the potential $\phi (z)$ as a product of
this plane wave with the $z$-dependent wave function $\psi (z) $ satisfying the Schr\"odinger-like equation ($c = 1$):
\begin{eqnarray}
\psi '' (z) + \frac{1}{\hbar^2}\left[\left(p_0 - e\phi (z)\right)^2 - \left(p^{2}_{\perp} + m^2\right)\right]\psi (z) = 0\,.\label{1.2}
\end{eqnarray}

We shall consider the smooth step potential of the Sauter type
\begin{eqnarray}
e\phi (z) = v \tanh kz\,,\label{1.3}
\end{eqnarray}
with $v ,k > 0$, where the sharp step potential is recovered in the limit $k\rightarrow\infty$ , while the limit $k\rightarrow 0$
with fixed $vk$ reproduces the linear potential due to a constant electric field.

In the potential~(\ref{1.3}), Eq.~(\ref{1.2}) can be solvable exactly~\cite{nikishov2}.
The solution describes the barrier scattering of a particle impinging the left with asymptotic boundary conditions
\begin{eqnarray}
\psi (z)\longrightarrow\left\{
\begin{array}{ccc}
A_1\,e^{ip_3^{ (-)}z/\hbar} + A_2\,e^{-ip_3^{ (-)}z/\hbar}&\mbox{,}&z\rightarrow -\infty\\&\\
B\,e^{ip_3^{ (+)}z/\hbar}&\mbox{,}&z\rightarrow +\infty
\end{array}\right.\,,\label{1.4}
\end{eqnarray}
where $A_1$, $A_2$ and $B$ are normalization constants. Let us introduce the initial and final values of
the particle energy far to the left and to the right of the potential~(\ref{1.3}):
\begin{eqnarray}
P_0^{(\mp)}\equiv\left.P_0 (z)\right|_{z\rightarrow\mp\infty} = p_0\pm v  ,
\label{1.5a}\end{eqnarray}
and the corresponding momenta
\begin{eqnarray}
p_3^{(\mp)}\equiv\left.p_3 (z)\right|_{z\rightarrow\mp\infty} = \sqrt{P_0^{(\mp)}{}^2 - (p^{2}_\perp + m^2)}\,.
\label{1.5}\end{eqnarray}
In the scattering process, the momenta $p_3^{(\mp)}$ in Eq.~(\ref{1.5}) are real thus restricting the asymptotic energies to $P_0^{(\mp)} > \sqrt{p^{2}_\perp + m^2}$.

To solve Eq.~(\ref{1.2}) with the asymptotic conditions~(\ref{1.4})
we set
\begin{eqnarray}
p_3 ^{(-)} \equiv 2\hbar k\mu\,,\quad
p_3 ^{(+)}\equiv  2\hbar k\nu\,,
\label{1.6}\end{eqnarray}
where
\begin{eqnarray}
\mu^2 - \nu^2 = \frac{vp_0  }{\hbar^2 k^2}\,.
\label{1.7}\end{eqnarray}
We further
replace  $z$  by the dimensionless variable
\begin{eqnarray}
\y = - \exp(-2kz)\,,
\label{1.8}\end{eqnarray}
running from $-\infty$ to $0$. As usual,
we extract
the asymptotic behavior at $z\rightarrow +\infty \,(\y\rightarrow 0)$
with the help of the substitution
\begin{eqnarray}
\psi (\y) = (-\y)^{-i\nu}\,f (\y)\,.
\label{1.9}\end{eqnarray}
This brings Eq.~(\ref{1.2})
to the following differential equation for $f (\y)$:
\begin{eqnarray}
\y f'' (\y) + (1-2i\nu) f' (\y) + \frac{v  }{k^2 \hbar^2}\left[\frac{v  }{(1-\y)^2} - \frac{p_0}{(1-\y)}\right]
f (\y) = 0 \,.
\label{1.10}\end{eqnarray}
The singularity at $\y=1$ suggests replacing
\begin{eqnarray}
f (\y) = (1 - \y)^{\delta}\,w (\y)\,,
\label{1.11}\end{eqnarray}
with
\begin{eqnarray}
\delta \equiv \frac{1}{2}+\bar  \delta \equiv
\frac{1}{2}\left( 1 + \sqrt{1 - \frac{4v^2}{k^2 \hbar^2}}\right)\,,
\label{1.12}\end{eqnarray}
leading to
a hypergeometric equation
for the function $w (\y)$:
\begin{eqnarray}
\y (1-\y) w'' (\y) + \left\{(1-2i\nu) - \left[(1-2i\nu) + 2\delta\right]\y\right\} w' (\y) - \left[\delta^2 -2i\nu\delta  + \frac{v p_0}{k^2 \hbar^2}\right] w (\y) = 0.
\label{1.13}\end{eqnarray}
The solution is the hypergeometric function
\begin{eqnarray}
w (\y) = F \left[\delta - i\left(\mu + \nu\right),\,\delta + i\left(\mu - \nu\right),\, 1 - 2i\nu\,;\,\y \,\right]\,
\label{1.14}\end{eqnarray}
up to some normalization factor.
For $z\rightarrow +\infty \,(\y\rightarrow 0)$, this function tends to $1$, and the solution of Eq.~(\ref{1.2}) contains only the
transmitted wave $\psi (z)\sim\exp [ip_3^{(+)}z/\hbar]$ satisfying the asymptotic condition in Eq.~(\ref{1.4}) with $B\sim 1$.
For $z\rightarrow -\infty \,(\y\rightarrow -\infty)$,
we find the asymptotic form of the function $\psi (z)$
in accordance with the asymptotic condition of Eq.~(\ref{1.4})
via the Kummer transformation of the hypergeometric function $w (\y)$,
with the coefficients
\begin{eqnarray}
A_1 &=& \frac{\Gamma (1-2i\nu)\,\Gamma (-2i\mu)}{\Gamma [\delta-i(\mu+\nu)]\,\Gamma [1-\delta-i(\mu+\nu])}\,,\\
A_2 &=& \frac{\Gamma (1-2i\nu)\,\Gamma (2i\mu)}{\Gamma [\delta+i(\mu-\nu)]\,\Gamma [1-\delta+i(\mu-\nu])}\,.
\label{1.15}\end{eqnarray}
From these we determine the reflection and transmission coefficients
for the Sauter potential:
\begin{eqnarray}
R\,=\,\frac{|A_2|^2}{|A_1|^2}\,,\quad\quad T\,=\,\frac{1}{|A_1|^2}\,.
\label{1.16}\end{eqnarray}
In our discussion of pair production, we shall focus mostly on the reflection coefficient
$R$.
To find a simple expression for it, we relate
the total kinetic energy $P_0^{(-)} - P_0^{(+)} = 2v  $
to the parameters $v,k$ of the Sauter potential~(\ref{1.3}) via Eq.~(\ref{1.12}).
For $4v^2/k^2 \hbar^2 < 1$, the square root
$\bar  \delta\equiv\sqrt{1 - 4v^2/k^2 \hbar^2}/2$ in Eq.~(\ref{1.12}) is real.
After substituting $A_1$ and $A_2$ from Eq.~(\ref{1.15}) into Eq.~(\ref{1.16}), the reflection coefficient takes the form
\begin{eqnarray}
R\,=\,\frac{\cos^2 \pi\bar\delta + \sinh^2 \pi(\mu-\nu)}{\cos^2 \pi\bar\delta + \sinh^2 \pi(\mu+\nu)}\,,
\label{1.17}\end{eqnarray}
where we have used the relation $\Gamma (1/2+z)\Gamma (1/2-z)=\pi/\cos\pi z$.

In the limit $k\rightarrow\infty$ where $\bar\delta\rightarrow 1/2$, Eq.~(\ref{1.17}) reproduces the well-known coefficient~\cite{caloger}:
\begin{eqnarray}
R\,= \,\left[\frac{p_3^{(-)} - p_3^{(+)}}{p_3^{(-)} + p_3^{(+)}}\right]^2\,
\label{1.18}\end{eqnarray}
for
the reflection of a relativistic particle off a potential step
\begin{eqnarray}
\phi (z) = a\left[\theta (z) - \theta (-z)\right] = \left\{
\begin{array}{ccc}
a&\mbox{,}&z > 0\\
\!\!- a&\mbox{,}&z < 0
\end{array}\right.\,,\label{1.19}
\end{eqnarray}
which arises from Eq.~(\ref{1.3}) in this limit.

For $4v^2/k^2 \hbar^2 > 1$, the square root $\bar\delta$ in Eq.~(\ref{1.12}) becomes purely imaginary,
$\bar\delta\equiv i\kappa$, where
\begin{equation}
\kappa\equiv\frac{1}{2}\sqrt{4v^2/k^2 \hbar^2 - 1}
\label{1.20a}\end{equation}
is real. Then the reflection coefficient in Eq.~(\ref{1.16}) with
$A_1$ and $A_2$ of Eq.~(\ref{1.15}) takes the form
\begin{eqnarray}
R\,=\,\frac{\cosh^2 \pi\kappa + \sinh^2 \pi(\mu-\nu)}{\cosh^2 \pi\kappa + \sinh^2 \pi(\mu+\nu)}\,,
\label{1.20}\end{eqnarray}
where we have used the relation $\Gamma (1/2+iy)\Gamma (1/2-iy)=\pi/\cosh\pi y$.

A special case of uniform electric field along the $z$-direction is
included here in the limit $k\rightarrow 0,\,v\rightarrow\infty$ with $vk =-eE=$ const. It corresponds to a linear potential
\begin{eqnarray}
\phi (z) = - Ez\,,\quad E<0\,,
\label{1.21}\end{eqnarray}
in Eq.~(\ref{1.3}). In this limit,
$\kappa\approx v/k\hbar$ becomes very large,
and $\mu\approx\nu\approx \kappa/2 - (p^{2}_{\perp} + m^2)/4\hbar e|E|$,
so that the reflection coefficient~(\ref{1.20a}) reduces to
\begin{eqnarray}
R\,=\,\left\{1 + \exp\left[ -\frac{\pi (p^{2}_{\perp} + m^2)}{\hbar (e|E|)}\right]\right\}^{-1}\,.
\label{1.22}\end{eqnarray}

We have assumed in this section that
the particles can pass through repulsive potential barrier with the
same (positive) sign of the initial and final energies $P_0^{(-)} = \left.P_0 (z)\right|_{z\rightarrow -\infty} = p_0 + v   >0$ and
$P_0^{(+)} = \left.P_0 (z)\right|_{z\rightarrow +\infty} = p_0 - v   >0$, respectively.
In the opposite situation, when the initial energy $P_0^{(-)} = p_0 + v   > 0$ is
positive but the final $P_0^{(+)} = p_0 - v   < 0$  becomes negative, the
famous {\em Klein paradox\/} arises. For a particle moving from
left to right this means that the region of large positive $z$ can only be
accessible to antiparticle.
Then a non-zero transmission coefficient must be present even for
a strong potential $\Delta v >2m$, where $\Delta v\equiv e\phi^{(+)} - e\phi^{(-)} = 2v  $ is the potential energy difference at infinity. As
has been explained in Refs.~\cite{hund,schwinger,hawking,sauter,caloger,hansen},
this happens for $p_0 + \ev   \geq m$ and $p_0 - \ev   \leq - m$, implying the spontaneous production of particle-antiparticle pairs with the total  energy difference $P_0^{(-)} - P_0^{(+)} = 2v  \geq 2m$.

\section{Pair production}
The pair production from the vacuum is now derived as follows. The average number of created pairs in the scattering process yields
the same result as the imaginary part of the effective action in Schwinger calculation~\cite{nikishov1,nikishov2}.
This means that the probability for vacuum to remain a vacuum under the influence of the external potential, i.e., the vacuum persistence probability, is related to the reflection coefficient by
\begin{eqnarray}
|\langle0|0\rangle|^2 = \prod_{p_0,\,\bf{p_{\perp}}} R_{p} = \exp\left(\sum_{p_0,\,\bf{p_{\perp}}}\ln R_{p}\right) \,,
\label{2.1}\end{eqnarray}
where the product and the sums are taken over all relevant quantum numbers $p_0$ and $\bf{p_{\perp}}$
of the created particles. Correspondingly, the pair production probability is
\begin{eqnarray}
P = 1 - \exp\left(\sum_{p_0,\,\bf{p_{\perp}}}\ln R_{p}\right)\approx - \sum_{p_0,\,\bf{p_{\perp}}}\ln R_{p} \,.\label{2.2}\end{eqnarray}
In a box-like volume $V_\perp T$ with $V_{\perp} = \int d^2 x_{\perp}= \int dx_1dx_2$ being the area of the potential step
transverse to the $z$-direction, and $T$ the total time, the sum over $p_0$ and $\bf{p_{\perp}}$ becomes an integral
\begin{eqnarray}
\sum_{p_0,\,\bf{p_{\perp}}}\ln R_{p} = V_{\perp}T\,\int\frac{d^2 p_{\perp}}{(2\pi\hbar)^2}\,\int\frac{d p_{0}}{(2\pi\hbar)}\,\ln R (p_0 ,\,p_{\perp})\,,
\label{2.3}\end{eqnarray}
where the reflection coefficient $R (p_0 ,\,p_{\perp})$ is
defined by Eq.~(\ref{1.17}) or (\ref{1.20}), and $p_{\perp}\equiv |\bf{p_{\perp}}|$.
The rotation invariance around the third axis reduces the integral $\int d^2 p_{\perp}$ to $\pi\int d p_{\perp}^2$.
The remaining integral over $p_0$ and $p_{\perp}^2$ is done over the Klein region
\begin{eqnarray}
P_0^{(-)} = p_0 + v   \geq \sqrt{p^{2}_\perp + m^2}\,,\quad\quad
P_0^{(+)} = p_0 - v   \leq - \sqrt{p^{2}_\perp + m^2}\,.
\label{2.4}\end{eqnarray}
The change of the sign of $P_0 (z)$ is necessary for a vacuum pair production by the Sauter potential~(\ref{1.3}) with $\ev > m$.

The pair production probability is now completely defined by Eqs.~(\ref{2.2})--(\ref{2.4}). The probability per unit area and unit time is
\begin{eqnarray}
w_\perp = - \frac{1}{2(2\pi)^2 \hbar^3 }\,\int^{(\ev ^2 -m^2)}_{0}\,d p_{\perp}^2
\,\int^{\ev - \sqrt{p^{2}_\perp + m^2}}_{- \ev   + \sqrt{p^{2}_\perp + m^2}}\,d p_{0}\,\ln R (p_0,\,p_{\perp}^2)\,,
\label{2.5}\end{eqnarray}
where the integration region in the $(p_{\perp}^2,p_0)$-plane is shown
in Fig.~1.
\begin{figure}[t]
\quad\quad\quad
\quad$\!\!\!        $
\quad
\quad\quad\quad\quad\quad
\quad\quad
\includegraphics[width=55mm,angle=0]{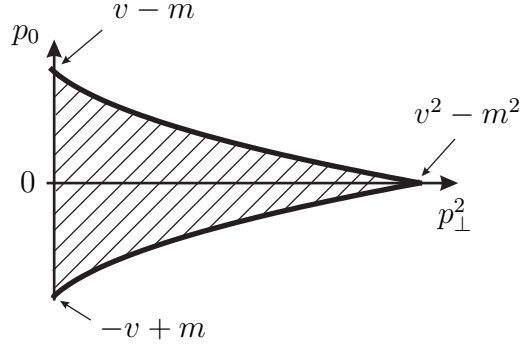}
\begin{picture}(0,10)
\put(-173,107){${p}_0$}
\put(-170,49){$0$}
\put(-12,37){${p}^2_\perp$}
\put(-135,115){$v-m$}
\put(-140,-6){$-v+m$}
\put(-21,75){$v^2-m^2$}
\end{picture}
\caption{\label{Fig1} In the $(p_{\perp}^2,p_0)$-plane, the integration covers the positive region
restricted by two intersecting parabolas $p_0 = \ev - \sqrt{p^{2}_\perp + m^2}$ and $p_0 = - \ev + \sqrt{p^{2}_\perp + m^2}$
with horizontal axes of symmetry above and below the $p_{\perp}^2$-axis for $\ev > m$.}
\end{figure}
For the actual calculation,
we interchange the order of integration in Eq.~(\ref{2.5}) to
\begin{eqnarray}
w_\perp\!\! =-\frac{1}{2(2\pi)^2\hbar^3 }
\left[\!\int^{0}_{-\ev + m}\!\!\!d p_{0}\int^{(p_0 + \ev  )^2 -m^2}_{0}\!\!\!\! d p_{\perp}^2+\int^{\ev - m}_{0}\!\!\!d p_{0}\int^{(p_0 - \ev)^2 -m^2}_{0}\!\!\!\!d p_{\perp}^2\right]\!\ln R (p_0,p_{\perp}^2).
\label{2.6}\end{eqnarray}
We now replace $p_0 \rightarrow - p_0$ in the first integral of Eq.~(\ref{2.6}) and
make use of the symmetry of the reflection coefficient $R (-p_0, p_{\perp}^2) = R (p_0, p_{\perp}^2)$
under the interchanging $\mu\leftrightarrow\nu$ in Eqs.~(\ref{1.17}) and (\ref{1.20}). The new integration region in the $(p_0, p_\perp^2)$-plane
is shown in Fig.~2. As a result, we obtain for the pair production rate per unit area in the Sauter potential~(\ref{1.3}) the integral
representation
\begin{eqnarray}
w_\perp =-\frac{1}{(2\pi)^2 \hbar^3 }\,
\int^{\ev  -m}_{0}\,d p_{0}\,\int^{(p_0 - \ev)^2 -m^2}_{0}\,d p_\perp^2 \,\ln R (p_0,p_{\perp}^2)\,.
\label{2.7}\end{eqnarray}

\comment{Most simply the reduced probability~(\ref{2.7}) can be calculated in the case of a constant electric field
where the reflection coefficient~(\ref{1.22}) is $p_0$-independent, whose logarithm has the expansion
\begin{eqnarray}
\ln R (p) = - \sum_{n=1}^{\infty}\,\frac{(-1)^{n+1}}{n}\,\exp\left[-\frac{(p_\perp^2 + m^2)}{\hbar (e|E|)}\pi n\right] \,.
\label{2.8}\end{eqnarray}
In Eq.~(\ref{2.7}) the integration is extended to whole positive infinite square in the $(p_0,\,p)$-plane.
Substituting Eq.~(\ref{2.8}) into Eq.~(\ref{2.7})
we must evaluate the energy integral. as $\int_{0}^{\infty}d p_0 = e|E|L/2$ yields the Schwinger result~\cite{schwinger}:
\begin{eqnarray}
w_\perp =L \frac{(e|E|)^2}{8\pi^3 \hbar^2 c}\sum_{n=1}^{\infty}\,\frac{(-1)^{n+1}}{n^2}\,\exp\left[-\frac{m^2 c^3}{\hbar (e|E|)}\pi n\right] \,,
\label{2.9}\end{eqnarray}
which amounts to the constant pair production per volume $w=w_\perp/ L$ found by Schwinger~\cite{schwinger}. We have reinserted the light velocity
$c$ for completeness. The same result was rederived by Nikishov~\cite{nikishov2} to demonstrate the equivalence of his and Schwinger's method,
and later by many other authors (see e.g.~\cite{kim} and references therein).
Apart from this special case of a constant electric field, no more exact results for the Sauter potential were found. In the purpose of this paper
we find the solution for an arbitrary potential step of the Sauter type (\ref{1.3}).}

\begin{figure}[h]
\quad\quad\quad\quad\quad\quad\quad\quad\quad\quad\quad
\includegraphics[width=60mm,angle=0]{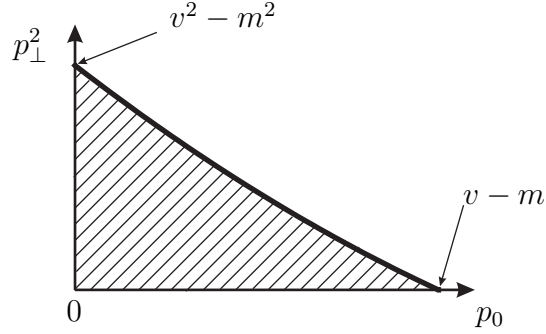}
\begin{picture}(0,12)
\put(-185,105){${p}^2_\perp$}
\put(-165,4){$0$}
\put(-10,4){${p}_0$}
\put(-15,48){$v-m$}
\put(-126,115){$v^2-m^2$}
\end{picture}
\caption{\label{Fig2} In the $(p_0,p_\perp^2)$-plane, the integration covers the region under the
left branch of the parabola \mbox{$p_\perp^2=(p_0 - \ev)^2 - m^2$} in the first quadrant for $\ev  >m$.}
\end{figure}

\section{Potential Step}
For a potential step of the Sauter type~(\ref{1.3}), the reflection coefficient as a function of $p_0$ and $p_{\perp}^2$
is given by Eq.~(\ref{1.18}). Its logarithm reads
\begin{eqnarray}
\ln R (p_0,p_{\perp}^2) = 2\ln\left[\frac{q_{+} (p_0,p_{\perp}^2) -  q_{-} (p_0,p_{\perp}^2)}
{q_{+} (p_0,p_{\perp}^2) +  q_{-} (p_0,p_{\perp}^2)}\right]\,,
\label{3.1}\end{eqnarray}
where $q_{\pm}(p_0,p_{\perp}^2)\equiv p_3^{(\mp)}=\sqrt{(p_0 \pm \ev  )^2 - (p_\perp^2 + m^2)}$ with $q_{+}\geq q_{-}$ for $p_0\geq 0$. As in Eq.~(\ref{1.7}), the functions $q_{\pm}(p_0,p_{\perp}^2)$ satisfy the constraint
\begin{eqnarray}
\left[q_{+} (p_0,p_{\perp}^2)\right]^2 - \left[q_{-} (p_0,p_{\perp}^2)\right]^2 = 4\ev p_0\,.
\label{3.2}\end{eqnarray}
The constraint suggests introducing a parameter $\theta$ so that
\begin{eqnarray}
q_{+} (p_0,p_{\perp}^2) = \sqrt{4\ev p_0}\,\cosh{\theta}\,,\quad
q_{-} (p_0,p_{\perp}^2) = \sqrt{4\ev p_0}\,\sinh{\theta}\,.
\label{3.3}\end{eqnarray}
This allows us to express Eq.~(\ref{3.1}) in terms of $\theta$ as
\begin{eqnarray}
\ln R (p_0,p_{\perp}^2) = - 4\theta\,,
\label{3.4}\end{eqnarray}
where
\begin{eqnarray}
{\theta} = {\theta}_{+}(p_0,p_{\perp}^2) = \ln\left[\frac{q_{+} (p_0,p_{\perp}^2) + q_{-} (p_0,p_{\perp}^2)}{\sqrt{4\ev p_0}}\right]\,.
\label{3.5}\end{eqnarray}
with $0\leq {\theta}\leq {\theta}_{+}(p_0, 0)$ for $0\leq p_\perp^2\leq (p_0 - \ev  )^2 -m^2$.
It is useful to eliminate the variable $p^2_\perp$ in favor of $\theta$ in the first integral of Eq.~(\ref{2.7}).

Alternatively we could have defined
\begin{eqnarray}
{\theta} = {\theta}_{-}(p_0,p_{\perp}^2) = \ln\left[\frac{q_{+} (p_0,p_{\perp}^2) - q_{-} (p_0,p_{\perp}^2)}{\sqrt{4\ev p_0}}\right]\,,
\label{3.6}\end{eqnarray}
with ${\theta}_{-}(p_0, 0)\leq {\theta}\leq 0$ for $0\leq p_\perp^2\leq (p_0 - \ev  )^2 -m^2$ where ${\theta}_{-}(p_0,p_{\perp}^2) = - {\theta}_{+}(p_0,p_{\perp}^2)$ due to Eq.~(\ref{3.2}). This definition corresponds to Eqs.~(\ref{3.3}) and (\ref{3.4}) with ${\theta}\rightarrow -{\theta}$.

The change of the measure in the first integral of Eq.~(\ref{2.7}) due to substituting $p^{2}_{\perp}\rightarrow\theta (p_0,p_{\perp}^2)$ is
\begin{eqnarray}
d p_\perp ^2= - 4\ev p_0\sinh \!2{\theta}\,d {\theta}\,.
\label{3.7}\end{eqnarray}
With Eqs.~(\ref{3.4}) and~(\ref{3.7}), the pair production rate~(\ref{2.7}) takes the form
\begin{eqnarray}
w_\perp = \frac{\ev  }{\pi^2 \hbar^3 }
\int^{\ev  -m}_{0} d p_{0}\,p_0\int^{{\theta}_{+}(p_{0},\,0)}_{0} d {\theta}\,4{\theta}\,\sinh 2{\theta}\,,\label{3.8}\end{eqnarray}
where the integration region in the $(p_0, {\theta})$-plane is shown in Fig.~3.
\begin{figure}[t]
\quad\quad\quad\quad\quad\quad\quad\quad\quad
\includegraphics[width=70mm,angle=0]{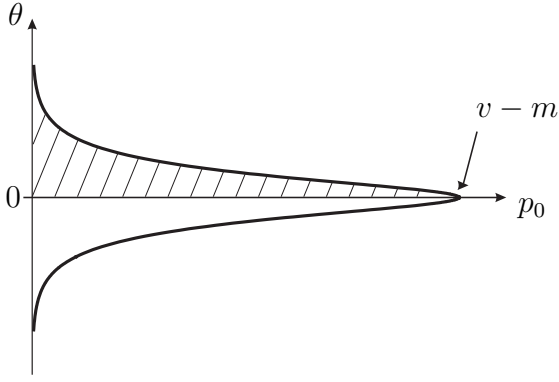}
\begin{picture}(0,10)
\put(-203,133){$\theta$}
\put(-204,64){$0$}
\put(-10,62){${p}_0$}
\put(-26,98){$v-m$}
\def\ssz{\scriptsize}
\end{picture}
\caption{\label{Fig3} The integration covers the upper region restricted by the logarithmic curve ${\theta} = {\theta}_{+}(p_0, 0)$ and the positive $p_0$-axis in the $(p_0, {\theta})$-plane.  Below the $p_0$-axis lies the alternative region restricted by the mirrored curve ${\theta} = {\theta}_{-}(p_0, 0)= - {\theta}_{+}(p_0, 0)$.}
\end{figure}
The integrals in Eq.~(\ref{3.8}) are now straightforward to do. The right-hand integral yields
\begin{eqnarray}
\int^{{\theta}_{+}(p_{0},\,0)}_{0}
d {\theta}\,4{\theta}\,\sinh 2{\theta}
= \frac{1}{2\ev p_0}\left[\left(q^2_{+} + q^2_{-}\right)
\ln\frac{q_{+} + q_{-}}{\sqrt{4\ev p_0}} - q_{+}\,q_{-}\right]_{p_{\perp}^2 = 0}.
\label{3.9}\end{eqnarray}
After this, the remaining integral in Eq.~(\ref{3.8}) becomes a combination of elliptic integrals via the substitution
$p_0 = (\ev  -m)t$ with $0\leq t \leq 1$. The last term in (\ref{3.9}) leads to
\begin{eqnarray}
\int^{\ev  -m}_{0}\,d p_{0}\,q_{+} (p_0, 0)\,q_{-} (p_0, 0) =
 \alpha_{+}\alpha_{-}^2\,I_2 (\alpha)\,,\label{3.10}\end{eqnarray}
where $I_2 (\alpha)$ is the dimensionless $t$-integral
\begin{eqnarray}
I_2 (\alpha) = \int^{1}_{0} d t\sqrt{(1 - t^2)(1 - \alpha^{2} t^2)}
= -\frac{1}{3\alpha^2}\left[(1 - \alpha^2)\,{\bf K}(\alpha) - (1 + \alpha^2){\bf E}(\alpha)\right]\,,
\label{3.10a}\end{eqnarray}
and ${\bf K}(\alpha)$, ${\bf E}(\alpha)$ are complete elliptic integrals of the first and second kind, respectively, with the
argument $\alpha \equiv\alpha_{-}/\alpha_{+}<1$, where $\alpha_{\pm} = \ev  \pm m$. With a little more effort we transform $p_0$-integral
over the first term in (\ref{3.9}) into a $t$-integral:
\begin{eqnarray}
\int^{\ev  -m}_{0}\!\!d p_{0}\left[q^2_{+}(p_0,0) + q^2_{-}(p_0,0)\right]\ln\left[\frac{q_{+}(p_0,0) + q_{-}(p_0,0)}{\sqrt{4\ev p_0}}\right]
= \alpha_{+}\alpha_{-}^2\,I_1 (\alpha)\,,\label{3.11}\end{eqnarray}
where $I_1 (\alpha)$ is the dimensionless integral
\begin{eqnarray}
I_1 (\alpha) = \int^{1}_{0} d t\,(1 + \alpha t^2)
\left\{2\ln\left[\sqrt{(1+t)(1 + \alpha t)} + \sqrt{(1-t)(1 - \alpha t)}\right] - \ln t -
\ln 2(1 + \alpha)\right\}\,.\label{3.11a}\end{eqnarray}
This can be expressed in terms of the elliptic integrals of the first and second kind $F(\varphi,1/\alpha)$ and $E(\varphi,1/\alpha)$
as follows:
\begin{eqnarray}
I_1 (\alpha) = \frac{2}{9\alpha}\left[(1 - \alpha)(4 + \alpha)\,F(\varphi, 1/\alpha) + \left(1 + \alpha (3 + \alpha)\right)
E (\varphi,1/\alpha)\right]\,,\label{3.12}\end{eqnarray}
with $\varphi = \arcsin\alpha$.

Finally, collecting all contributions in Eq.~(\ref{3.8}) yields the pair production rate per area
\begin{eqnarray}
w_\perp &=& \frac{\ev ^3}{3\pi^2 \hbar^3}\left(1 + \frac{m}{\ev  }\right)^3
\left\{\frac{1}{2}\left[(1 - \alpha^2)\,{\bf K}(\alpha) - (1 + \alpha^2){\bf E}(\alpha)\right]
\right.\nonumber\\
&+&\left.\frac{\alpha}{3}\left[(1 - \alpha)(4 + \alpha)\,F
\left(\varphi, \frac{1}{\alpha}\right) + \left(1 +3 \alpha  + \alpha^2\right)
E \left(\varphi,\frac{1}{\alpha}\right)\right]\right\}
\,.\label{3.13}\end{eqnarray}

\section{Sauter potential}
We employ now Eq.~(\ref{2.7}) to compute the pair production probability for the Sauter potential~(\ref{1.3})
where the reflection coefficient is defined by Eqs.~(\ref{1.17}) and (\ref{1.20}) for all values of the parameters $v$
and $k$. In order to illustrate the calculation, we specify these, for example, as $(2\ev  )^2/k^2 \hbar^2 > 1$.
In this case, the parameter $k$ defines the inverse width of the electric field, whereas the
parameter $v$ governs its size $|E|=vk/e$, whose maximum is $|E_{c}|\equiv {m^2 c^3}/{e \hbar}\simeq 1.3\times 10^{18}\,\,$V/m.
The limit $k\rightarrow 0$ with fixed $vk$ reproduces the linear potential due to a constant electric field.

The reflection coefficient of the Sauter potential with $4v^2/k^2 \hbar^2 > 1$ is given by  Eq.~(\ref{1.20}). An equivalent form of this is
\begin{eqnarray}
R = \frac{\cosh\pi(\mu - \nu + \kappa)\,\cosh\pi(\mu - \nu - \kappa)}{\cosh\pi(\mu + \nu + \kappa)\,\cosh\pi(\mu + \nu - \kappa)}\,,
\label{4.1}\end{eqnarray}
where $\mu = \mu (p_0,p_{\perp}^2)$ and $\nu = \nu (p_0,p_{\perp}^2)$
are the functions of $p_0$ and $p_{\perp}^2$ defined by Eqs.~(\ref{1.5a}), (\ref{1.5}) and (\ref{1.6}) with the constraint~(\ref{1.7}),
while $\kappa =\sqrt{4v^2 -k^2\hbar^2}/2k\hbar$ is a constant. Taking logarithms of  Eq.~(\ref{4.1}) leads to the expansion
\begin{eqnarray}
\ln R  = - 4\sum_{n=1}^{\infty}\,\frac{(-1)^{n+1}}{n}\,\cosh 2\pi n\kappa\sinh 2\pi n\mu\sinh 2\pi n\nu\,.
\label{4.2}\end{eqnarray}
The right hand side is found by replacing each logarithm of the hyperbolic functions
as $\ln (2\cosh x) = x + \ln (1 + e^{-2x}) = x + \sum_{n=1}^\infty (-1)^{n+1} e^{-2nx}/ n$, and combining all sums.

With Eq.~(\ref{4.2}), the pair production rate per area~(\ref{2.7}) takes the form
\begin{eqnarray}
w_\perp = \frac{1}{\pi^2 \hbar^3}\sum_{n=1}^{\infty}\,\frac{(-1)^{n+1}}{n}\,\cosh(2\pi n\kappa)\,J^{(n)}\,,
\label{4.3a}\end{eqnarray}
where $J^{(n)}$ are the integrals
\begin{eqnarray}
J^{(n)}\! = \int^{\ev  -m}_{0}\! d p_{0}\,I^{(n)}(p_0)\equiv
\int^{\ev  -m}_{0}\! d p_{0} \int^{(p_0 - \ev  )^2 -m^2}_{0}\!d p_\perp^2\,
\sinh 2\pi n\mu (p_0,p_{\perp}^2)\,\sinh 2\pi n\nu (p_0,p_{\perp}^2)\,.
\label{4.3}\end{eqnarray}
The region of integration is shown in Fig.~2.

A physically more instructive quantity than the production rate~(\ref{2.7}) can be obtained by dividing $w_\perp$
by the width of the potential step. For the Sauter potential the width is defined by the ratio
\begin{equation}
L = \int_{-\infty}^\infty dz\, E^2(z)/E^2_{\rm max} = 4/3k.
\label{4.3b}\end{equation}
Thus we obtain the pair creation rate per volume of nonzero field
\begin{eqnarray}
w = w_\perp/L = (3k/4) w_\perp.
\label{2.7a}\end{eqnarray}

We perform the $p_\perp^2$-integration in Eq.~(\ref{4.3}) by changing,
for each $n$ separately, from the variable $p_{\perp}^2$ to
the dimensionless one  $\theta$ defined by
\begin{equation}
p_{\perp}^2\rightarrow {\theta}\equiv \left[2\pi n\nu (p_0,p_{\perp}^2)\right]^2
\label{4.4a}\end{equation}
with $0\leq{\theta}\leq \bar\theta^{(n)}(p_0)$, where
\begin{eqnarray}
\bar\theta^{(n)}(p_0)\equiv \left[2\pi n\nu (p_0,0)\right]^2 = (n\pi/k\hbar)^{2}\left[\left(p_0 - v \right)^2 -m^2\right]\,.
\label{4.4b}\end{eqnarray}
For each $n$, Eq.~(\ref{4.4a}) yields $2\pi n\nu (p_0,p_{\perp}^2) = \sqrt{\theta}$ and $d p_{\perp}^{2} = - (k\hbar/n\pi)^2\,d \theta$.
Let us also introduce the dimensionless functions
\begin{equation}
{\theta ^{(n)}(p_0)}\equiv (2\pi n/k\hbar)^2{\ev p_0}\,,
\label{4.4c}\end{equation}
so that $2\pi n\mu (p_0,p_{\perp}^2) = \sqrt{\theta^{(n)}(p_0) + \theta}$ due to Eq.~(\ref{1.7}).
Then $p_\perp^2$-integrals take the form
\begin{eqnarray}
\!\!\!\!\!\!\!\!\!\!\!\!I^{(n)}(p_0)&=&
\int^{(p_0 - \ev  )^2 -m^2}_{0}\!d p^2_\perp\,\sinh 2\pi n\mu (p_0,p_{\perp}^2)\,\sinh 2\pi n\nu (p_0,p_{\perp}^2)\nonumber\\
\!\!&=&\frac{(k\hbar)^2}{2(n\pi)^2}\!
\int^{\bar\theta^{(n)}(p_0)}_{0}\!d{\theta}\!\left\{\!\cosh\!\!\left(\!\sqrt{\theta^{(n)}(p_0) + {\theta}} + \sqrt{{\theta}}\right)\!\! - \cosh\!\!\left(\!\sqrt{\theta^{(n)}(p_0)+ {\theta}} - \sqrt{{\theta}}\right)\!\right\}\!.
\label{4.4}\end{eqnarray}
The two terms in Eq.~(\ref{4.4}) can now be combined into a single integral as follows. We substitute
\mbox{$t=\sqrt{\theta^{(n)}(p_0) + {\theta}} + \sqrt{{\theta}}$} with
$\sqrt{\theta^{(n)}(p_0)}\leq t\leq  \theta^{(n)}_{+}(p_0)$ in the first term, and \mbox{$t=\sqrt{\theta^{(n)}(p_0) + {\theta}} - \sqrt{{\theta}}$}
with $\theta^{(n)}_{-}(p_0)\leq t\leq \sqrt{\theta^{(n)}(p_0)}$ in the second, where
\begin{eqnarray}
\theta^{(n)}_{\pm}(p_0)\equiv (2\pi n)[\mu (p_0, 0)\pm \nu (p_0, 0)].
\label{4.4d}\end{eqnarray}
Noting that $\sqrt{{\theta}}=\pm [t/2 - \theta^{(n)}(p_0)/2t]$ in the first and the second substitution,
respectively, while $d{\theta}=(t/2)\{1-[\theta^{(n)}(p_0)]^2/t^4\}\,d t$ in both cases, we obtain
\begin{eqnarray}
I^{(n)}(p_0) = \left(\frac{k\,\hbar}{2\pi n}\right)^2\int^{\theta^{(n)}_{+}(p_0)}_{\theta^{(n)}_{-}(p_0)} d t\,t\left\{1 - \frac{[\theta^{(n)}(p_0)]^2}{t^4}\right\}\,\cosh t\,.
\label{4.5}\end{eqnarray}
Evaluating this integral yields
\begin{eqnarray}
I^{(n)}(p_0) = I^{(n)}_{+}(p_0) + I^{(n)}_{-}(p_0)\,,
\label{4.6}\end{eqnarray}
with
\begin{eqnarray}
I^{(n)}_{\pm}(p_0)
\!=\!\!\left(\!\frac{k\,\hbar}{2\pi n}\!\right)^{2}\!\!\left\{\!\pm\!\left[\theta^{(n)}_{\pm}\!\sinh\!\theta^{(n)}_{\pm}\!- \cosh\!\theta^{(n)}_{\pm}\right]\!
\mp\frac{\left[\theta^{(n)}_{+}\theta^{(n)}_{-}\right]^2}{2}\!\!\left[{\rm Chi}
\,\theta^{(n)}_{\pm}\!- \frac{\sinh\!\theta^{(n)}_{\pm}}{\theta^{(n)}_{\pm}} -
\frac{\cosh\!\theta^{(n)}_{\pm}}{\left[\theta^{(n)}_{\pm}\right]^2}\right]\!\right\}\!,
\label{4.7}\end{eqnarray}
where
$\theta^{(n)}_{\pm}$ is short for $\theta^{(n)}_{\pm}(p_0)$,
${\rm Chi}\, \theta^{(n)}_{\pm}$ are the hyperbolic cosine integrals, and the last two terms represent the leading terms
in their asymptotic expansions for large arguments $\theta^{(n)}_{\pm}$.

Having obtained $I^{(n)}(p_0)= I^{(n)}(\theta^{(n)}_{+}(p_0)\,,\theta^{(n)}_{-}(p_0))$, we are left in Eq.~(\ref{4.3}) with the sum
\begin{eqnarray}
J^{(n)} = \int^{\ev  -m}_{0} d p_{0}\,I^{(n)}(p_0) = J^{(n)}_{+} + J^{(n)}_{-}\,,
\label{4.8}\end{eqnarray}
of the integrals over the rather lengthy functions~(\ref{4.7}):
\begin{eqnarray}
J^{(n)}_{\pm} = \int^{\ev  - m}_{0} d p_{0}\,I^{(n)}_{\pm}(p_0)\,.
\label{4.9}\end{eqnarray}
However, this sum can be combined into a single integral by subjecting $J^{(n)}_{\pm}$ in Eq.~(\ref{4.9})
to a change of variables $p_0\rightarrow\xi$ provided that we define the new dimensionless integration variable $\xi$ as follows.

In order to transform the integral $J^{(n)}_{+}$, we introduce the dimensionless variable
\begin{eqnarray}
\xi (p_0) = \mu (p_0, 0) + \nu (p_0, 0)\,,
\label{4.10}\end{eqnarray}
with $\sqrt{\ev (\ev -m)}/\hbar k\leq \xi\leq \sqrt{\ev ^2 - m^2}/\hbar k$ for \mbox{$0\leq p_0 \leq (\ev  -m)$} and $\ev  >m$,
where $\mu (p_0, 0)\equiv\sqrt{(p_0 + \ev  )^2 - m^2}/2\hbar k$ and $\nu (p_0, 0)\equiv\sqrt{(p_0 - \ev  )^2 - m^2}/2\hbar k$.
Noting that $m^2\leq\left[\ev ^2 - \kxi^2\right]$ within these limits, we resolve Eq.~(\ref{4.10}) in terms of $p_0$ as
\begin{eqnarray}
p_{0} (\xi) =\hbar k\xi\left\{1 - \frac{m^2}{\left[\ev ^2 - \kxi^2\right]}\right\}^{1/2}\,,
\label{4.11}\end{eqnarray}
with positive $\xi$ due to $p_0\geq 0$. To determine $\theta^{(n)}_{\pm}(p_0)$ of Eq.~(\ref{4.4d}) in terms of $\xi$ by means of Eq.~(\ref{4.11}), we find first for a given $\xi$ the positive square roots
\begin{eqnarray}
\sqrt{(p_0 \pm \ev  )^2 - m^2} =\hbar  k\xi\pm v  \left\{1 - \frac{m^2}{\left[\ev ^2 - \kxi^2\right]}\right\}^{1/2}\geq 0\,.
\label{4.12}\end{eqnarray}
Combining these yields
\begin{eqnarray}
\mu (p_0, 0) + \nu (p_0, 0) = \xi\,,\quad \mu (p_0, 0) - \nu (p_0, 0)=
\frac{\ev}{\hbar k}\left\{1 - \frac{m^2}{\left[\ev ^2 - \kxi^2\right]}\right\}^{1/2}\,.
\label{4.13}\end{eqnarray}
From Eq.~(\ref{4.13}) we obtain, finally, the functions  $\theta^{(n)}_{\pm}(\xi)\equiv \theta^{(n)}_{\pm}(p_0(\xi))$ to be substituted
instead of $\theta^{(n)}_{\pm}(p_0)$ into the first integral $J^{(n)}_{+}$ as
\begin{eqnarray}
\theta^{(n)}_{+}(\xi)={2\pi n}\xi\,,\quad \theta^{(n)}_{-}(\xi)=\frac{2\pi n}{\hbar k}\ev \left\{1 - \frac{m^2}{\left[\ev ^2 - \kxi^2\right]}\right\}^{1/2}\,.
\label{4.14}\end{eqnarray}
Note that the inequality $m^2\leq\left[\ev ^2 - \kxi^2\right]$ ensures the
positivity of expressions under the square roots in Eqs.~(\ref{4.11})--(\ref{4.14}).

In  order to treat the second integral $J^{(n)}_{-}$ in Eq.~(\ref{4.9}), we define a new integration variable $\xi$
similar to Eq.~(\ref{4.10}):
\begin{eqnarray}
\xi (p_0) = \mu (p_0,0) - \nu (p_0,0)\,,
\label{4.15}\end{eqnarray}
with $0 \leq \xi\leq \sqrt{\ev (\ev -m)}/\hbar k$ for \mbox{$0\leq p_0 \leq (\ev  - m)$} and $\ev > m$, where
$\mu (p_0, 0)\equiv\sqrt{(p_0 + \ev  )^2 - m^2}/2\hbar k$ and $\nu (p_0, 0)\equiv\sqrt{(p_0 - \ev  )^2 - m^2}/2\hbar k$.
Since $m^2\leq v m\leq \left[\ev ^2 - \kxi^2\right]$ in these limits, we solve Eq.~(\ref{4.15}) in terms of $p_0$
in the same way as in Eq.~(\ref{4.11}), leading again to
\begin{eqnarray}
p_{0} (\xi ) =\hbar k\xi\left\{1 - \frac{m^2}{\left[\ev ^2 - \kxi^2\right]}\right\}^{1/2}\,,
\label{4.16}\end{eqnarray}
with positive $\xi$ due to $p_0\geq 0$. To determine $\theta^{(n)}_{\pm}(p_0)$
in terms of $\xi$ by means of Eq.~(\ref{4.16}), we find now for a given $\xi$ the positive square roots
\begin{eqnarray}
\sqrt{(p_0 \pm \ev  )^2 - m^2} = \pm \hbar k\xi  + v\left\{1 - \frac{m^2}{\left[\ev ^2 - \kxi ^2\right]}\right\}^{1/2}\geq 0\,.
\label{4.17}\end{eqnarray}
It follows from Eq.~(\ref{4.17}) that
\begin{eqnarray}
\mu (p_0,0) - \nu (p_0,0) = \xi\,,\quad \mu (p_0,0) + \nu (p_0,0)=
\frac{\ev}{\hbar k}\left\{1 - \frac{m^2}{\left[\ev ^2 - \kxi ^2\right]}\right\}^{1/2}\,.
\label{4.18}\end{eqnarray}
This yields $\theta^{(n)}_{\pm}(\xi)$ which replaces
$\theta^{(n)}_{\pm}(p_0)$ in the second integral $J^{(n)}_{-}$ as follows:
\begin{eqnarray}
\theta^{(n)}_{+}(\xi )=\frac{2\pi n}{\hbar k}\ev \left\{1 - \frac{m^2}{\left[\ev ^2 - \kxi ^2\right]}\right\}^{1/2}\,\,\,,
\quad \theta^{(n)}_{-}(\xi )={2\pi n}\xi \,.
\label{4.19}\end{eqnarray}
Again, the inequality $m^2\leq\left[\ev ^2 - \kxi ^2\right]$ ensures
the positiveness of
the expressions under the square roots in Eqs.~(\ref{4.16})--(\ref{4.19}).

We now go from the integration variable $p_0$ to $\xi$ in Eq.~(\ref{4.9}) for $J^{(n)}_{+}$ and $J^{(n)}_{-}$ separately.
Substituting these in Eq.~(\ref{4.8}) yields
\begin{eqnarray}
\!\!\!\!J^{(n)}&=&\frac{\ev ^2}{2}\left(\frac{2\pi n}{\hbar k}\right)^2
\int^{{\bar\xi }}_{0}\!d \xi\,\frac{d p_{0} (\xi)}{d \xi}\,p^{2}_{0}(\xi)\,
{\rm Chi}\left({2\pi n}\,\xi\right)\nonumber\\
\!\!\!\!&-&\frac{\ev ^2}{2}\frac{1}{\hbar^2 k^2}
\int^{{\bar\xi }}_{0}\!d \xi\,\frac{d p_{0}
(\xi)}{d \xi}\,\frac{p^{2}_{0}(\xi)}{\xi^2}\,\left[
\cosh\left({2\pi n}\,\xi\right)
+2\pi n\xi\,\sinh\left(2\pi n\,\xi\right)\right]\nonumber\\
\!\!\!\!&+&\frac{\hbar^2 k^2}{(2\pi n)^2}
\int^{{\bar\xi }}_{0}\!d \xi\,\frac{d p_{0} (\xi)}{d \xi}\,
\left[\cosh\left({2\pi n}\,\xi\right) -{2\pi n} \xi\,\sinh\left({2\pi n}\,\xi\right)\right]\,,
\label{4.20}\end{eqnarray}
where
\begin{eqnarray}
{\bar\xi }\equiv\sqrt{\ev ^2 -m^2}/\hbar k,
\label{4.20a}\end{eqnarray}
and the function $p_0 (\xi)$ is given by Eq.~(\ref{4.11}) [or (\ref{4.16})].
Equation~(\ref{4.20}) can be simplified by partial integration thanks to
the vanishing of the function $p_0 (\xi)$ at
the endpoints of integration. This yields
\begin{eqnarray}
J^{(n)}= (\hbar k)^3\int^{{\bar\xi }}_{0}\!d \xi\, f (\xi)\,\cosh \left({2\pi n}\,\xi\right)\,,
\label{4.21}\end{eqnarray}
where the function $f (\xi)$ has the form
\begin{eqnarray}
f (\xi)\equiv \xi\,p_{0}(\xi)/\hbar k- \ev ^2\,p^{3}_{0}(\xi)/3\,(\hbar k)^{5}\xi^3\,.
\label{4.22}\end{eqnarray}
After inserting $p_{0}(\xi)$ from Eq.~(\ref{4.11}) [or (\ref{4.16})],
this reads explicitly
\begin{eqnarray}
f (\xi) = \xi^2\!\left[\frac{{\bar\xi }^2 - \xi^2}{{\bar\xi }^2 - \xi^2+
\left(m/\hbar k\right)^2}\right]^{1/2}\! - \frac{{\bar\xi }^2 + \left(m/\hbar k\right)^2}{3}\!
\left[\frac{{\bar\xi }^2 -\xi^2}{{\bar\xi }^2 - \xi^2 + \left(m/\hbar k\right)^2}\right]^{3/2}\! = f(-\xi).
\label{4.23}\end{eqnarray}
Note that the integral over this function vanishes:
$\int^{\bar \xi}_{0}\!d \xi\, f (\xi)=0 $, so that $J^{(0)}=0$.

Alternatively, we may introduce the variable $\eta \equiv \xi/{\bar\xi }$ to arrive at the integral
\begin{eqnarray}
J^{(n)}= \sqrt{v^2 - m^2}^3\!\int^{1}_{0}\!d \eta\, g (\eta)\,\cosh \left({2\pi n{\bar\xi }}\,\eta\right)\,,
\label{4.24}\end{eqnarray}
with
\begin{eqnarray}
g (\eta)=f(\xi (\eta))/{\bar\xi }^2 = \eta^2\,\left[\frac{1 - \eta^2}{1- \eta^2 + \gamma^2}\right]^{1/2}
\!\!- \frac{1+ \gamma^2}{3}\left[\frac{1-\eta^2}{1 - \eta^2+ \gamma^2}\right]^{3/2}\!\! = g(-\eta)\,,
\label{4.25}\end{eqnarray}
where $\gamma\equiv m/\hbar k{\bar\xi } = m/\sqrt{v^2-m^2}$.
The integrals $J^{(n)}$ are all functions of $v,m$, and $k$.

\section{Pair Production Rates}
Let us first check our final expression~(\ref{4.3a}) with (\ref{4.24}) by going
to the limit of a constant electric field $k\rightarrow 0,\,v\rightarrow\infty$ with $vk = e|E|$,
where the exact result is known.
In this limit, the parameter $\gamma$ becomes small and can be neglected, and the
integrals~(\ref{4.24}) become  approximately
\begin{eqnarray}
J^{(n)}\simeq \frac{\ev ^3}{(\pi {\bar\xi })n}\left[\frac{\sinh(2\pi{\bar\xi }n)}{3} - \frac{\cosh(2\pi{\bar\xi }n)}{(2\pi{\bar\xi })n}
+ \frac{\sinh(2\pi{\bar\xi }n)}{(2\pi{\bar\xi })^2 n^2}\right] \,.
\label{5.1}\end{eqnarray}
Inserting these into Eq.~(\ref{4.3a}) leads to the following hyperbolic sums
\begin{eqnarray}
\sum_{n=1}^{\infty}\,\frac{(-1)^{n+1}}{n^{\nu}}\times\left\{
\begin{array}{ccc}
\cosh(n \lambda_{\pm})&\mbox{}&\\&\\
\sinh(n \lambda_{\pm})&\mbox{}&
\end{array}\!\!\!\!\!\!\!\!\!\!\!\!\right\} = -\frac{1}{2}{\rm Li}_{\nu}(-e^{\lambda_{\pm}})
\mp \frac{1}{2}{\rm Li}_{\nu}(-e^{-\lambda_{\pm}})\,,
\label{5.2}\end{eqnarray}
with $\nu = 2,3,4$, where $\lambda_{\pm}\equiv 2\pi (\kappa \pm {\bar\xi })$ and ${\rm Li}_{\nu}(z)$ are the polylogarithm functions
\begin{eqnarray}
{\rm Li}_{\nu}(z)\equiv\sum_{n=1}^{\infty}\,\frac{z^n}{n^{\nu}}\,,\quad \nu = 2,3,4\,.
\label{5.2a}\end{eqnarray}
Note that the constant field limit corresponds to large arguments in Eq.~(\ref{5.2}), since $\kappa\simeq \ev/\hbar k$, ${\bar\xi }\simeq\ev/\hbar k - m^2/2\ev \hbar k  \simeq\kappa - \rho/2\pi$, where
\begin{eqnarray}
\rho  =\pi m^2/\hbar e|E|\,.
\label{5.2b}\end{eqnarray}
We must exploit therefore the analytic continuation of the polylogarithm functions defined by the series~(\ref{5.2a}) into
the region $|z|>1$. By taking advantage of the formula \cite{form}:
\begin{eqnarray}
{\rm Li}_\nu(- z) + e^{i\pi \nu}{\rm Li}_\nu(- z^{-1})= \frac{(2\pi)^ \nu }{\Gamma(\nu)}
e^{i\pi\nu/2}\zeta\left(1-\nu,\frac{1}{2}+\frac{\log (z)}{2\pi i}\right)\,,
\label{5.2c}\end{eqnarray}
where $\zeta(\nu ,q)$ is the Hurwitz zeta function
\begin{equation}
\zeta(\nu ,q) = \sum_{n=0}^\infty \frac{1}{(z+q)^\nu}\,,
\label{5.2d}\end{equation}
we bring the right-hand side of (\ref{5.2}) to  the form
\begin{eqnarray}
-\frac{1}{2}\frac{(2\pi)^ \nu }{\Gamma(\nu)}e^{i\pi\nu/2}\zeta\left(1-\nu,\frac{1}{2}+\frac{\lambda_{\pm}}{2\pi i}\right)
+ e^{i\pi\nu}{\rm Li}_{\nu}(-e^{-\lambda_{\pm}}).
\label{5.3a}\end{eqnarray}
For $\nu = 2,3,4$ this reads explicitly,
\begin{eqnarray}
\sum_{n=1}^{\infty}\,\frac{(-1)^{n+1}}{n^{2}}\,\sinh(n \lambda_{\pm})&=&\frac{\pi^2}{12}+\frac{\lambda_{\pm}^2}{4}+ {\rm Li}_{2}(-e^{-\lambda_{\pm}})\,,\nonumber\\
\sum_{n=1}^{\infty}\,\frac{(-1)^{n+1}}{n^{3}}\,\cosh(n \lambda_{\pm})&=&\frac{\pi^2}{12}\lambda_{\pm}+\frac{\lambda_{\pm}^3}{12}- {\rm Li}_{3}(-e^{-\lambda_{\pm}})\,,\label{5.3}\\
\sum_{n=1}^{\infty}\,\frac{(-1)^{n+1}}{n^{4}}\,\sinh(n \lambda_{\pm})&=&\frac{7\pi^4}{720}+\frac{\pi^2}{24}\lambda_{\pm}^2 +\frac{\lambda_{\pm}^4}{48}+ {\rm Li}_{4}(-e^{-\lambda_{\pm}})\nonumber
\,.\end{eqnarray}
Substituting Eq.~(\ref{4.3a}) with Eqs.~(\ref{5.1}) and~(\ref{5.3}) into Eq.~(\ref{2.7a}), we obtain the approximate pair
production rate per nonzero field volume
\begin{eqnarray}
w&\!\!\!\!=\!\!\!\!&w _\perp/L\simeq\frac{3k}{4\pi^2 \hbar^3}\frac{\ev^3}{(\pi{\bar\xi })}
\!\left\{\frac{2\pi^2}{3}(\kappa{\bar\xi })+\frac{1}{6}{\rm Li}_{2}(-e^{-\lambda_{+}})-\frac{1}{6}{\rm Li}_{2}(-e^{-\lambda_{-}})\right.\nonumber\\
&\!\!\!\!-\!\!\!\!&\left.\frac{1}{(2\pi{\bar\xi })}\!\left[\frac{\pi^3}{6}\kappa + \frac{2\pi^3}{3}\kappa\!\left(\kappa^2 + 3{\bar\xi }^{2} \right)-\frac{1}{2}{\rm Li}_{3}(-e^{-\lambda_{+}})-\frac{1}{2}{\rm Li}_{3}(-e^{-\lambda_{-}})\right]\right.\nonumber\\
&\!\!\!\!+\!\!\!\!&\left.\frac{1}{(2\pi{\bar\xi })^2}\!\left[\frac{\pi^4}{3}(\kappa{\bar\xi })+
\frac{4\pi^4}{3}(\kappa{\bar\xi })\!\!\left(
\kappa^2 + {\bar\xi }^{2} \right)+\frac{1}{2}{\rm Li}_{4}(-e^{-\lambda_{+}})-
\frac{1}{2}{\rm Li}_{4}(-e^{-\lambda_{-}})\right]\!\right\}\!.
\label{5.4a}\end{eqnarray}
We now take the constant-field limit $k\rightarrow 0,\,v\rightarrow\infty$ at $vk = e|E|$
fixed, where $\lambda_{-} = \rho$ remains finite, while $\lambda_{+} = 4\pi\kappa - \rho  \simeq 4\pi\kappa\rightarrow 4\pi v/\hbar k\simeq 4\pi eE/k^2\hbar $ tends to infinity, so that
${\rm Li}_{\nu}(-e^{-\lambda_{+}})$ with $\nu = 2,3,4$ vanishes.
Moreover, the polylogarithm functions ${\rm Li}_{\nu}(-e^{-\lambda_{-}})$
with $\nu = 3,4$ do not contribute because of vanishing prefactors.
All divergent terms cancel each other.
Thus we obtain the pair production rate per nonzero field volume,
which for constant field is the total volume:
\begin{eqnarray}
w=\frac{w _\perp}L\rightarrow - \frac{3}{4}\frac{(e|E|)^2}{6\pi^3 \hbar^2c}\,{\rm Li}_{2}(-e^{- \rho })\,.
\label{5.4}\end{eqnarray}
Here we have inserted $L=4/3k$ from Eq.~(\ref{4.3b}).
The division by $L$ is essential for getting a finite result in the constant-field limit.
For completeness, we have reinserted in the final expression the light velocity $c$ to verify the
complete agreement with the result of Heisenberg and Euler~\cite{heeu},
Schwinger~\cite{schwinger}, Nikishov~\cite{nikishov2}, and many others (see e.g. \cite{kim} and references therein).

For arbitrary $k$, the integral~(\ref{4.21}) cannot be evaluated in closed
analytic form. In order to obtain an approximate
rate formula we insert $J^{(n)}$ from Eq.~(\ref{4.21})
into Eq.~(\ref{4.3a}) and interchange the order
of summation and integration to find the expansion
\begin{eqnarray}
w_\perp = \frac{k^3}{2\pi^2  }\,\int^{\bar \xi}_{0}\!d \xi\, f (\xi)
\,\sum_{n=1}^{\infty}\,\frac{(-1)^{n+1}}{n}\,\left\{\cosh \left[ 2\pi n (\xi - \kappa)\right]
+ \cosh \left[ 2\pi n (\xi + \kappa)\right] \right\}\,.
\label{5.5}\end{eqnarray}
The integral is simplified with the help of the summation formula
\begin{eqnarray}
\sum_{n=1}^{\infty}\,\frac{(-1)^{n+1}}{n}\,\cosh(n x) =
\frac{1}{2}\,\left[\ln\left(1 + e^{-x}\right) + \ln\left(1 + e^{x}\right)\right]\,,
\label{5.6}\end{eqnarray}
which permits us to bring the general rate to the form
\begin{eqnarray}
w_\perp = \frac{k^3}{2\pi^2}\,\int^{\bar \xi}_{0}\!d \xi\, f (\xi)
\left\{\ln\left[1 + e^{-2\pi (\xi - \kappa)}\right] + \ln\left[1 + e^{2\pi (\xi + \kappa)}\right]\right\}\,,
\label{5.7a}\end{eqnarray}
or, by the symmetry of the function $f(\xi) = f(-\xi)$, to the more symmetric form
\begin{eqnarray}
w_\perp = \frac{k^3}{2\pi^2}\,\int^{\bar \xi}_{-\bar \xi}\!d \xi\, f (\xi)
\,\ln\left[1 + e^{-2\pi (\xi - \kappa)}\right]\,.
\label{5.7}\end{eqnarray}
Finally, integrating this by parts, we find
\begin{eqnarray}
w_\perp = \frac{k^3}{\pi }\,
\int^{\bar \xi}_{-\bar \xi}\!d \xi\,g (\xi)\,\frac{1}{e^{2\pi (\xi - \kappa)} +1}\,,
\label{5.8}\end{eqnarray}
where the function $g (\xi)$ vanishes on both ends. Explicitly, it reads
\begin{eqnarray}
g (\xi) = - \frac{\xi}{3}\,
\frac{\left(\bar \xi^2 - \xi^2\right)^{3/2}}{\left(\bar \xi^2 - \xi^2 + m^2/\hbar^2 k^2\right)^{1/2}} = - g (-\xi)\,.
\label{5.9}\end{eqnarray}
Remarkably, the second function under the integral~(\ref{5.8}) resembles a Fermi distribution.
Indeed, we are going to show that the calculation of the integral~(\ref{5.8}) can be done by
a method familiar to low-temperature expansions in statistical physics~\cite{kv}.
The condition necessary for pair production $v = e|E|/k > m$ implies that the parameter $k$ lies
in the interval $0 < k <\epsilon m/\hbar$, where
\begin{equation}
\epsilon\equiv |E|/|E_{c}|
\label{5.10}\end{equation}
and $E_{c}\equiv m^2/e\hbar$ (in natural units with $c=1$)
is the so-called critical field for which the work over two Compton wavelengths $2\hbar /m$ can produce the energy $2m$ of a pair.
At the upper end $k = \epsilon m/\hbar$ of the above interval,
the rate~(\ref{5.8}) vanishes, since  $\bar \xi$ becomes zero.
For the calculation of the exact pair production
rate~(\ref{2.7a}) from Eq.~(\ref{5.7}),
we introduce the dimensionless parameter $\tilde k\equiv\hbar k/\epsilon m$, where
$0 < \tilde k < 1$, to rewrite Eq.~(\ref{5.8}) in terms of the dimensionless variable $\tilde \xi\equiv\xi \hbar k /\ev$.
This brings the production rate~(\ref{2.7a}) to the form
\begin{eqnarray}
w = - \frac{(e|E|)^2}{4\pi\hbar^2}\,\frac{1}{{(\epsilon{\tilde k}^2})^2}\,
\int^{\hat\xi}_{-\hat\xi}
\!d \tilde \xi\, \tilde g (\tilde \xi)\,
\frac{1}{e^{2\pi(\tilde\xi - \tilde\kappa)/\epsilon\tilde k{}^2} +1 }\,,
\label{5.11}\end{eqnarray}
where
\begin{equation}
{\hat\xi}\equiv\sqrt{1 - {\tilde k}^2}\,, ~~~~
\tilde\kappa\equiv\sqrt{1-(\epsilon\tilde k{}^2/2)^2}\,,
\label{5.12}\end{equation}
and the dimensionless function $\tilde g (\tilde \xi)$ reads
\begin{eqnarray}
\tilde g (\tilde \xi) = \tilde\xi\,\frac{({\hat \xi}^2 - {\tilde \xi}^2)^{3/2}}{\left(1 - {\tilde \xi}^2\right)^{1/2}}\,.
\label{5.13}\end{eqnarray}
We expand this function into a power series
\begin{eqnarray}
\tilde g (\tilde \xi) = \sum_{n=0}^\infty {\tilde g}_{2n+1} (\hat\xi) \tilde \xi^{2n+1} \,,
\label{5.14}\end{eqnarray}
with the coefficients
\begin{eqnarray}
{\tilde g}_1 (\hat\xi) =\hat\xi^3,
~~{\tilde g}_3 (\hat\xi) = - \hat\xi (3 - \hat\xi^2)/2,
~~{\tilde g}_5 (\hat\xi) = 3(1 -\hat\xi^2)^{2}/8\hat \xi\,,~~\dots\,.
\label{5.15}\end{eqnarray}
Substituting the expansion~(\ref{5.14}) back into Eq.~(\ref{5.11}), we encounter the odd-moment integrals of the
Fermi distribution
\begin{eqnarray}
M_{2n+1}(\hat \xi)\equiv \int^{\hat\xi}_{- \hat \xi}\!d \tilde\xi\,
\frac{{\tilde\xi}^{2n+1}}{e^{2\pi(\tilde\xi - \tilde\kappa)/\epsilon\tilde k{}^2}+1}\,,\,\,\,n\geq 0\,.
\label{5.16}\end{eqnarray}
These can all be found exactly. Performing the integrals yields a binomial expansion
\begin{eqnarray}
\!M_{2n+1} (\hat \xi)\!=\!\!\sum_{m=1}^{2n+2}\,(-1)^{m+1}\,\frac{\Gamma (2n+2)\hat \xi^{2n-m+2}(\epsilon\tilde k^2)^{m}}{(2\pi)^{m}\Gamma (2n-m+3)}\,\overline{\rm Li}_{m}(\hat \xi)\,.
\label{5.17}\end{eqnarray}
Here $\overline{\rm Li}_{m}(\hat \xi)$ are the linear combinations of the polylogarithm functions
\begin{eqnarray}
\overline{\rm Li}_{m}(\hat \xi)\equiv{\rm Li}_{m}(- {\tilde\lambda}^{(+)} (\hat \xi)) + (-1)^{m-1}\,{\rm Li}_{m}(-{\tilde\lambda}^{(-)}(\hat \xi))\,,
\label{5.18}\end{eqnarray}
with arguments
\begin{eqnarray}
{\tilde\lambda}^{(+)} (\hat \xi)\equiv e^{\tilde\rho (\hat \xi)}\,,\quad
{\tilde\lambda}^{(-)} (\hat \xi)\equiv{\tilde\lambda}^{(+)} (-\hat \xi)=
e^{\tilde\rho (-\hat \xi)}\,,
\label{5.19}\end{eqnarray}
where
\begin{eqnarray}
\tilde\rho(\hat \xi)\equiv 2\pi(\hat \xi - \tilde\kappa)/\epsilon\tilde k^2\,.
\label{5.20}\end{eqnarray}
The exact production rate of Eq.~(\ref{5.11}) becomes now the sum
\begin{eqnarray}
w = -\frac{(e|E|)^2}{4\pi \hbar^2}\,\frac{1}{{(\epsilon{\tilde k}^2})^2}\,\sum_{n=0}^\infty {\tilde g}_{2n+1} (\hat\xi)\,M_{2n+1} (\hat\xi)\,.
\label{5.21}\end{eqnarray}
By making use of Eqs.~(\ref{5.17}), we rewrite this as an expansion over the polylogarithm functions
\begin{eqnarray}
w = -\frac{(e|E|)^2}{4\pi \hbar^2}\,\sum_{m=1}^\infty c_{m} (\hat\xi)\,(\epsilon\tilde k^{2})^{m-2}\,\overline{\rm Li}_{m}(\hat\xi)\,,
\label{5.22}\end{eqnarray}
where the coefficients $c_{m} (\hat\xi)$ are polynomials of $\hat\xi$
\begin{eqnarray}
c_{m} (\hat\xi) = \sum_{n=0}^\infty \frac{(-1)^{m+1}\Gamma (2n+2)}{(2\pi)^{m}\Gamma (2n-m+3)}\,
{\tilde g}_{2n+1}(\hat\xi)\,\hat\xi^{2n-m+2} \,.
\label{5.23}\end{eqnarray}
Together with Eq.~(\ref{5.14}) these read explicitly,
\begin{eqnarray}
c_{1} (\hat\xi) &=& (5\hat\xi^4/256\pi)(-1 - 4\hat\xi^2/5 - 6\hat\xi^4/5 - 4\hat\xi^6 + 7\hat\xi^8 + \cdots)\,,\nonumber\\
c_{2} (\hat\xi) &=& (\hat\xi^3/512\pi^2)(125 + 84\hat\xi^2 + 102\hat\xi^4 + 260\hat\xi^6 - 315\hat\xi^8 + \cdots)\,,\nonumber\\
c_{3} (\hat\xi) &=& (3\hat\xi^2/128\pi^3)(15 - 10\hat\xi^2 - 32\hat\xi^4 - 110\hat\xi^6 + 105\hat\xi^8 + \cdots)\,,\label{5.24}\\
c_{4} (\hat\xi) &=& (3\hat\xi/256\pi^4)(- 205 - 70\hat\xi^2 + 132\hat\xi^4 + 910\hat\xi^6 - 735\hat\xi^8 + \cdots)\,,\nonumber\\
c_{5} (\hat\xi) &=& (1/256\pi^5)(1347 + 1296\hat\xi^2 - 18\hat\xi^4 - 9240\hat\xi^6 + 6615\hat\xi^8 + \cdots)\,,\nonumber\\
& \vdots&.\nonumber\end{eqnarray}

The series expansion given by Eq.~(\ref{5.22}) converges well for small $\tilde k$.
Here the parameter $\tilde\rho(\hat\xi)$ in Eq.~(\ref{5.20}) becomes
$\tilde\rho(\hat\xi)\simeq -\pi/\epsilon + \pi (\epsilon{\tilde k}^2)/4 + \cdots$,
where the first term is equal to $- \rho$ of Eq.~(\ref{5.2b}), and the polylogarithm functions
${\rm Li}_{m}(- {\tilde\lambda}^{(+)} (\hat \xi))$  with $m >2$
will be suppressed by powers of $\tilde k$. The parameter $\tilde\rho(-\hat \xi)\equiv -2\pi(\hat \xi+\tilde\kappa)/\epsilon\tilde k^2$
tends to minus infinity, so that the polylogarithm functions ${\rm Li}_{m}(- {\tilde\lambda}^{(-)} (\hat \xi))$ for all $m$  yield the exponentially small contributions. By means of Eq.~(\ref{5.12}), the coefficients~(\ref{5.24}) are the polynomials of small $\tilde k$
\begin{eqnarray}
c_{1} (\tilde k)&=& (1/\pi)({-\tilde k}^2/4 + 5{\tilde k}^4/4 - 43{\tilde k}^6/16 + \cdots)\,,\nonumber\\
c_{2} (\tilde k)&=& (1/\pi^2)(1/2 - 3{\tilde k}^2/8 - 27{\tilde k}^4/8 + 589{\tilde k}^6/64 + \cdots)\,,\nonumber\\
c_{3} (\tilde k)&=& (1/\pi^3)(-3/4 + 3{\tilde k}^2/8 + 189{\tilde k}^4/16 - 1011{\tilde k}^6/32 + \cdots)\,,\label{5.25}\\
c_{4} (\tilde k)&=& (1/\pi^4)(3/8 - 2655{\tilde k}^4/64 + 6555{\tilde k}^6/64 + \cdots)\,,\nonumber\\
c_{5} (\tilde k)&=& (1/\pi^5)(279{\tilde k}^4/2 - 19155{\tilde k}^6/64 + \cdots)\,,\nonumber\\
& \vdots&.\nonumber\end{eqnarray}
Finally, this yields the probability rate~(\ref{5.22}) as a series expansion in powers of small $\tilde k$
\begin{eqnarray}
w = &-& \frac{(e|E|)^2}{4\pi\hbar^2}\left\{\left[-\frac{1}{4\pi\epsilon}\,{\rm Li}_{1}(-e^{-\tilde\rho }) + \frac{1}{2\pi^2}\,{\rm Li}_{2}(-e^{-\tilde\rho })\right]\right.\nonumber\\
&+&\left. \tilde k^2\,\left[\frac{5}{4\pi\epsilon}\,{\rm Li}_{1}(-e^{-\tilde\rho }) - \frac{3}{8\pi^2}\,{\rm Li}_{2}(-e^{-\tilde\rho })
- \frac{3\epsilon}{4\pi^3}\,{\rm Li}_{3}(-e^{-\tilde\rho })\right] + \cdots\right\}\,,
\label{5.26}\end{eqnarray}
where the leading term is already an excellent approximation. Note the coincidence of the second term in the first brackets
with the probability rate~(\ref{5.4}) for a constant-field limit $k\rightarrow 0$.

\section{Conclusion}
We have calculated an exact expression for the production rate
of charged scalar particle-antiparticle pairs from the vacuum by the Sauter potential.
For an arbitrary potential barrier, the
rate was related to the scattering amplitude on the barrier,
and expressed
as an energy-momentum integral over the logarithm of the reflection coefficient.
For the Sauter potential, we
have evaluated this integral  and checked the result by
recovering the known limits of a sharp step potential and of a uniform electric field.

~\\
{Acknowledgement}:\\[2mm]
The authors are grateful for many discussions with
Remo Ruffini and She-Sheng Xue.

\end{document}